\documentclass[prl,aps,twocolumn,superscriptaddress,floatfix,showpacs]{revtex4}

\newcommand{\xp}{x^{\prime}}
\newcommand{\wfx}{\psi(x)}
\newcommand{\wfxp}{\psi(x^{\prime})}
\newcommand{\triple}{\mathrm{C}\equiv \mathrm{C}}
\newcommand{\single}{\mathrm{C}-\mathrm{C}}
\newcommand{\ch}{\mathrm{C}-\mathrm{H}}
\newcommand{\form}{\mathrm{C}_N \mathrm{H}_2}
\newcommand{\formmin}{\mathrm{C}_N \mathrm{H}_2^{-}}

\usepackage{graphicx,amssymb,amsmath}

\begin{document}

\title{Charge carrier solvation and large polaron formation on a polymer chain revealed in model \textit{ab initio} computations}

\author{M.~L.~Mayo}
\affiliation{Department of Physics, The University of Texas at
Dallas, P. O. Box 830688, FO23, Richardson, Texas 75083, USA}
\author{Yu.~N.~Gartstein}
\affiliation{Department of Physics, The University of Texas at
Dallas, P. O. Box 830688, FO23, Richardson, Texas 75083, USA}

\begin{abstract}
When an excess charge carrier is added to a
semiconducting polymer chain, it is well known that the
carrier may self-trap into a polaronic state accompanied by a
bond length adjustment pattern. A different
mechanism of self-localization is the solvation of  charge
carriers expected to take place when the polymer chain is immersed
in polar media such as common solvents. We use state-of-the-art \textit{ab initio} computations in conjunction with the Polarizable Continuum Model to unequivocally demonstrate solvation-induced self-consistent charge localization into large-radius one-dimensional (1D) polarons on long $\form$ carbon chains with the polyynic structure. Within the framework used, the solvation results in a much more pronounced charge localization. We believe this mechanism of polaron formation to be of relevance for various 1D semiconductors in polar environments.
\end{abstract}


\maketitle

A variety of one-dimensional (1D) wide-band semiconductor (SC) structures: conjugated polymers, nanowires, and nanotubes, attract a great deal of attention. The nature and properties of excess charge carriers on these structures are fundamental for various applications. For our discussion, one distinguishes between nearly free band states of excess carriers and self-localized polaronic \cite{alexmott} states. Self-trapping and the formation of 1D polarons can occur owing to the (strong) interaction of the electronic subsystem with another subsystem such as, e.g., displacements of the underlying atomic lattice, which is the mechanism that has been extensively explored for 1D electron-phonon systems. Here we are concerned with a different, much less studied implementation of the polaronic effect that should take place for 1D SCs immersed in 3D polar media, the situation particularly common for applications involving fundamental redox processes in polar solvents. In this case of what could be called charge carrier solvation, the long-range Coulomb interaction is expected \cite{conwell_1,gartstein_1,gartstein_3} to lead to the formation of a localized electronic state on a SC structure surrounded by a self-consistent pattern of the sluggish (orientational) polarization of the solvent. The main goal that we pursue and achieve in this paper is to substantiate the validity of this picture at the level of \textit{ab initio} computations, that is, with realistic many-electron effects taken into account. To our knowledge, this is a first \textit{ab initio} demonstration of the \textit{solvation-induced} self-consistent localization of the excess charge resulting in large-radius 1D polarons.

The type of systems we consider is an interesting model realization in the general context of quantum particles interacting with dissipative environments \cite{weiss99,YNGmob}. The confinement of the electron motion to 1D and its structural separation from the 3D polarizable medium clearly distinguishes such systems from the well-known 3D cases of polarons in polar SCs \cite{polarons1,polarons2,CTbook} and solvated electrons in polar liquids \cite{CTbook,ferra91}. A combination of several factors here creates conditions that may be especially favorable for the formation of 1D continuum adiabatic polarons; these are small effective band masses of the carriers (e.g., $\sim 0.1$ of free electron mass), enhanced Coulomb effects in 1D \cite{haugbook}, and the slow response of the solvent polarization.

The self-trapping of an excess charge carrier (an electron or a hole) is easily described in a \textit{single-particle} picture within a standard 1D continuum adiabatic framework. It corresponds to the localized ground-state solution $\wfx$ of the non-linear Schr\"{o}dinger equation:
$$
-\frac{\hbar^2}{2m}\frac{\partial^2 \wfx}{\partial x^2} \,-\,
\int \! d\xp V(x-\xp)\, |\wfxp|^2 \,\wfx = E\wfx,
$$
where $x$ is the coordinate along the structure aixs, $m$ the intrinsic effective mass of the carrier, and $V(x)$ the effective self-interaction mediated by another subsystem. In the case of a short-range electron-phonon mediation, the self-interaction can be taken local: $V(x)=g\,\delta(x)$ leading then to the exact result $\wfx \propto \mathrm{sech}(gmx/2\hbar^2)$ for a continuum 1D polaron widely known after the
pioneering contributions of Rashba \cite{rashba_LP} and Holstein \cite{holstein_LP}. In the case of the long-range polarization interaction, the effective $V(x)$ behaves as $1/|x|$ at large distances, while the short-range behavior depends on specific system details such as, e.g., the actual transverse charge density distribution and the geometry of the dielectric screening. Numerical studies of this case \cite{gartstein_1,gartstein_3} indicate that the binding energy of the resulting polarons could reach a substantial fraction, roughly one-third, of the binding energy of the corresponding Wannier-Mott excitons (primary optical excitations in many 1D SCs).

While illuminating the basic physics of the effect, the single-particle description has the drawback of not \textit{explicitly} including  valence band electrons, whose reorganization may substantially affect the outcomes. Such a reorganization can be important even in non-interacting electron models -- a nice illustration was given in Ref.~\cite{FVB} that analyzed how polarons of a two-band Peierls dielectric model evolve into single-particle Holstein polarons in the limit of the ``frozen valence band'' approximation.  With realistic Coulomb interactions in place, the role of valence electrons in the formation of the relevant self-consistent potentials would only increase. As applied to electron-phonon polarons in conjugated polymers (CPs), various \textit{ab initio} computational schemes have been found useful to clarify the role of electron-electron interactions on such polarons and their optical properties (we refer the reader to a recent account \cite{BredasPolOptics} for a discussion and multiple references). In what follows we use the Gaussian 03-implemented state-of-the-art DFT \textit{ab initio} calculations in conjunction with the Polarizable Continuum Model
(PCM) \cite{g03,leach_1} to study the polaron formation in  conjugated linear systems as \textit{mediated by the polarization interaction} with a surrounding polar solvent.

Since the computational demand increases substantially in DFT-PCM calculations,  we have selected structurally simple systems of hydrogen-terminated polyynic linear carbon chains $\form$ (even $N$ between 20 and 100) for our demonstration. A polyynic structure exhibits in its ground state an alternating pattern of triple $\triple$ and single $\single$ bonds and is semiconducting. It should be noted that, while playing the role of a prototypical example in our study, polyynic chains continue to be the subject of much attention in their own right \cite{yang_kertesz_2,schaefer_1,gladysz_2}.

The results presented in this paper have been derived by
employing DFT computations with the B3LYP hybrid exchange-correlation
functional known to include local and non-local correlation effects. The 6-311++G(\textit{d,p}) all electron basis set was chosen due to its richness and ability to represent charged systems. We note that our ``in vacuum'' calculations have been verified to compare well against previously published \textit{ab initio} results on neutral \cite{yang_kertesz_2} and charged \cite{schaefer_1} polyynic systems both in terms of energetics and optimized bond lengths. The DFT-PCM (``in solvent'') calculations used water as a polar solvent with its \textit{default} parameters in Gaussian 03.

\begin{figure}
\centering
\includegraphics[scale=1.0]{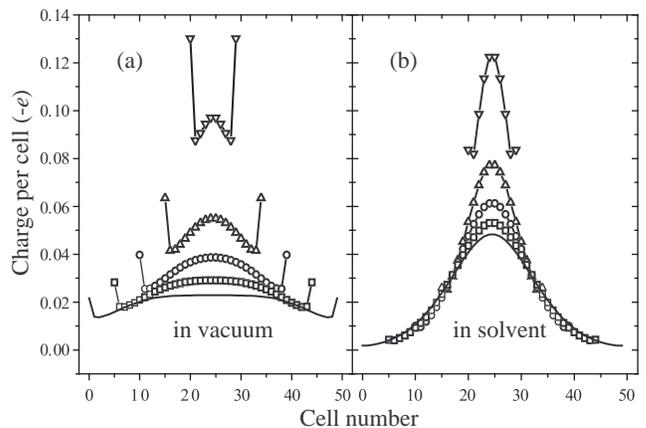}
\caption{Excess charge density spatial distribution on charged $\formmin$ chains with  $N=20$, 40, 60, 80 and 100 -- top to bottom curves, respectively, in their central parts. All chains feature the same pattern of fixed bond lengths corresponding to RG1 geometry described in text. Panel (a) displays results for chains in vacuum, panel (b) for chains in the solvent environment.}
\label{fig:ch_diff_r0}
\end{figure}

\begin{figure}
\centering
\includegraphics[scale=1.0]{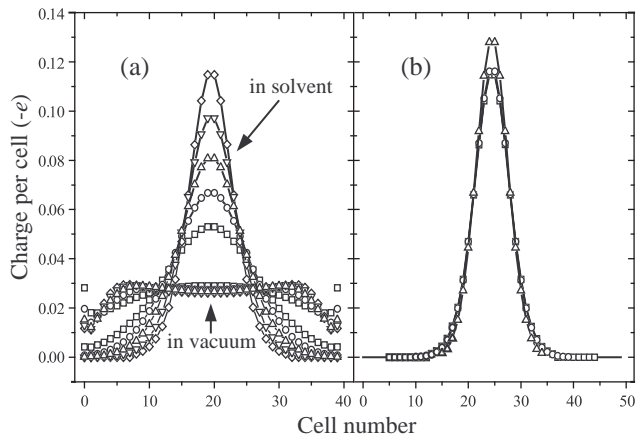}
\caption{Excess charge density spatial distribution on charged $\formmin$ chains. (a) N=80. The evolution of the charge distribution upon stretching the single bonds by 0.1 {\AA} \ successive increments from 1.326 {\AA} \ to 1.726 {\AA} \ -- bottom to top in a group of 5 clearly distinct curves shown by different connected symbols and denoted ``in solvent''. The same - but unconnected - data symbols are used to show the results for chains in vacuum, those distributions largely overlap. (b) Comparison of distributions for chains in the solvent with $N=40$, 60, 80 (different data symbols) and 100 (solid line) for the geometry RG2 featuring the longest single bonds.}\label{fig:ch_diff_c80h2_vac_soln}
\end{figure}

\begin{figure}
\centering
\includegraphics[scale=0.6]{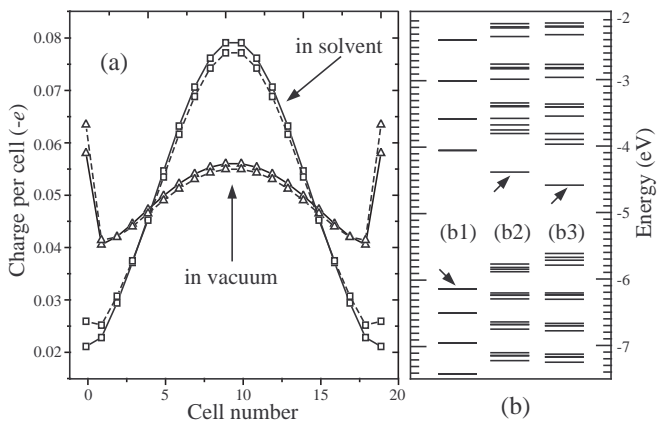}
\caption{Comparison of the solvation and bond-length adjustment effects for C$_{40}$H$_2$. (a) Excess charge distributions in vaccum ($\vartriangle$) and in the solvent ($\square$) for RG1 geometry (dashed lines) and for the fully optimized bond length patterns (solid lines). (b) The corresponding MO energy levels in the solvent for: (b1) Optimized neutral chain, (b2) Charged chain in RG1 geometry, (b3) Optimized charged chain. The arrows show HOMO levels.}
\label{fig:c40h2_opt_noopt}
\end{figure}

A pertinent quantity for us is the equilibrium \textit{spatial distribution} of the excess charge over the 1D SC, which is naturally related to atomic charges in outputs of \textit{ab initio} calculations. Importantly, these charges represent the effect due to \textit{all} electrons in the system. It is well known \cite{leach_1} that calculations of the atomic charges from many-electron wave functions are dependent on the basis set transformations and sometimes lead to spurious results. Two procedures, due to Mulliken and L\"{o}wdin, are used particularly frequently in such calculations \cite{leach_1}. We exercised both and established that, while they may lead to different raw atomic charges for individual cases, the results become remarkably close when calculated for unit-cell-based charge \textit{differences} between charged and neutral chains. The unit cells are defined here to consist of pairs of neighboring carbons, with the chain end cells including the hydrogen atomic charges. It is these \text{stable} charge differences -- that is, excess charge -- per cell that are displayed in Figs.~\ref{fig:ch_diff_r0}, \ref{fig:ch_diff_c80h2_vac_soln} and \ref{fig:c40h2_opt_noopt}(a) using L\"{o}wdin numbers. We studied both anionic (an extra electron) and cationic (an extra hole) chains. Their self-localization behavior has been found very similar. Accordingly we restrict our discussion here to the anionic cases, with the excess charge in the figures measured in units of $(-e)$, where $e$ is the magnitude of the fundamental charge.

To be able to clearly separate the effect of the solvation on excess charge localization from the effect of atomic displacements, we start by exploring systems with rigid geometries, that is, with fixed prescribed atomic positions. Our first geometry, referred to as RG1, features the following fixed lengths: 1.235~\AA \ for the triple bonds, 1.326~\AA \ for the single, and 1.063~\AA \ for the $\ch$ bonds. These lengths have been retrieved from our complete optimization of the neutral $N=80$ chain in vacuum and are therefore expected to be relatively close to optimal values. We now use this geometry to perform DFT calculations for a set of  $\form$ and $\formmin$ samples with various $N$ to find the excess charge density distribution both in vacuum and in the solvent environment. The derived results are compared in Fig.~\ref{fig:ch_diff_r0} whose (a) and (b) panels display clearly distinct trends. Barring the end effects, vacuum results in panel (a) are reminiscent of the ``particle in a box'' behavior: as the chain length increases, the excess charge is distributed more and more uniformly over the whole chain. In a sharp contrast, for chains in the solvent, panel (b), the excess charge exhibits much more localized distributions around central parts of the chains. As the chain length increases, those distributions show a tendency towards convergence, albeit not completely achieved within the range of lengths studied.

To further the understanding of the observed behavior, we now attempt to decrease the spatial extent of charge localization. In single-particle models, the size of the polaron  is determined by the balance of the gain in the potential energy of the localized carrier and the loss in its kinetic energy. Decreasing the latter leads to a diminished delocalization propensity and is expected to shorten the polaron. This should be achievable with an increased intrinsic effective mass, or with a narrower electron bandwidth. To imitate the effect, we chose to \textit{artificially} stretch all single bonds in our system, thereby making $\triple$ ``dimers'' more and more separated from each other. Figure \ref{fig:ch_diff_c80h2_vac_soln}(a) illustrates a \textit{dramatic} difference in the response of the excess charge distribution to successive single-bond stretchings for chains in vacuum and for chains in solvent. Whereas charge density in vacuum is hardly affected in this process, preserving nearly uniform distributions over the whole chain, for chains in solvent, each successive stretching step results in smaller and smaller localization regions. We refer to the fixed geometry with the longest-stretched single bonds of 1.726 \AA \ as RG2. As the localization size in this geometry appears to be well within the chain length, we compare the excess charge distributions for several chain lengths in Fig.~\ref{fig:ch_diff_c80h2_vac_soln}(b). The distributions for $N=60$, 80, and 100 practically coincide confirming that they correspond to a fully converged self-consistent pattern of excess charge localization achieved \textit{entirely} due to the interaction with the surrounding solvent.
\begin{figure}
\centering
\includegraphics[scale=0.52]{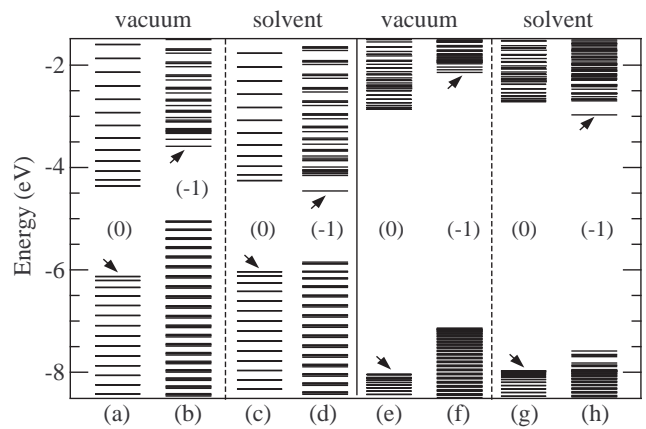}
\caption{Molecular orbital energy levels for neutral (0) and charged (-1) C$_{100}$H$_2$ chains with rigid RG1 (a-d) and RG2 (e-h) geometries both in vacuum and in the solvent. The arrows indicate positions of HOMO levels.}
\label{MOlevels}
\end{figure}

A comparison of the corresponding molecular orbital (MO) energy level structures is available in Fig.~\ref{MOlevels}. One immediately notices a much wider (but inconsequential for the effect we consider) band gap in the RG2 geometry. The narrowing of the bandwidth is, however, relevant and easily detected via a much larger density of states (panels (e) vs (a) and (g) vs (c)). Disregarding ``trivial'' (the electrostatic potential with respect to infinity) overall shifts of the levels for the charged chains in vacuum, the important difference with the charged chains in the solvent is that the latter case exhibits local highest occupied (HOMO) levels clearly separated from other states: compare panel (d) to (b) and, especially, panel (h) to (f). This is the behavior expected from single-particle polaron models. One, however, does not observe distinctly separate unoccupied local levels appearing in such models with long-range Coulomb potentials \cite{gartstein_1}. Also, a reorganization of valence band electron states is apparent in the band gap upon the formation of the polaron, especially well seen in geometry RG2, panel (h). This is suggestive of a more complex electronic structure of the polaron as affected by many-electron interactions.

To better benchmark the localization due to the solvation, we now turn to bond-length optimization results derived for the $N=40$ system. Figure \ref{fig:c40h2_opt_noopt}(a) pictures modifications of the excess charge spatial distribution that take place upon changes from the rigid RG1 geometry to fully optimized bond length patterns both in vacuum and in the solvent. It is evident that \textit{for the charge distribution}, the effect of the lattice relaxation is much less significant than the effect of the solvation. (The same conclusion has been drawn in our studies of $N=80$ chains.) Figure \ref{fig:c40h2_opt_noopt}(b) shows how local HOMO and intra-gap valence levels are affected by the pure solvation and by the combined action of the solvation and lattice displacements. Interestingly, here the effects due to each of the mechanisms appear similar in magnitude despite a much more pronounced effect of the solvation on charge density localization. This is again likely indicative of the intricacies of many-electron interactions and the screening effects.  A more complete understanding of similarities and differences of these mechanisms calls for further studies at various computational levels. The picture of Fig.~\ref{fig:c40h2_opt_noopt}(b) suggests that comparative spectroscopy of electronic transitions to and from local HOMO levels should be a helpful experimental tool.

The spatial extent of demonstrated solvation-induced 1D polarons is seen to be on the order of tens of Angstr\"{o}ms. We, however, do not feel to be able to reliably evaluate the polaron binding energy except to estimate it at $\sim 0.1$ eV based on comparisons of LUMO levels with the computed electron affinities. Our consideration has been limited to chains with idealized straight structures; various defects including spatial conformations of the chains can lead to the formation of bound polarons with possibly larger binding energies.

While providing a clear conceptual demonstration, the present \textit{ab initio} framework may overestimate the effects of the solvation on self-localization as it does not take a proper account of the frequency dependence of the solvent dielectric function \cite{polarons1,CTbook,gartstein_1,gartstein_3}. This issue could possibly be addressed within time-dependent DFT-PCM computational schemes \cite{SolvTDepend}. Another current development, on the scaling description of solvent effects \cite{SolvScaling}, may also help in studies of large polaron formation. Given the general character of the solvation-induced self-localization with the subsequent decrease of charge carrier mobility \cite{conwell_1,CBdrag,YNGmob}, as well as its expected significance for fundamental redox processes on various 1D SC structures, further
studies of this effect seem quite interesting and warranted.

We gratefully acknowledge financial support from the Collaborative U. T. Dallas -
SPRING Research and Nanotechnology Program.

\bibliography{abinitio_ref}

\end{document}